%% file: 0_sample-manuscript.tex
\documentclass[manuscript]{acmart}

\usepackage{graphicx, hyperref}
\usepackage{xcolor}
\input{1_figures}

\newcommand{\ib}{institutional betrayal }
\newcommand{\IB}{Institutional betrayal }

\AtBeginDocument{%
  }

\setcopyright{acmlicensed}
\copyrightyear{2025}
\acmYear{2025}
\acmDOI{XXXXXXX.XXXXXXX}

\acmISBN{978-1-4503-XXXX-X/18/06}

\begin{document}

\title[Platforms as Crime Scene, Judge, and Jury]{Platforms as Crime Scene, Judge, and Jury: How Victim-Survivors of Non-Consensual Intimate Imagery Report Abuse Online}

\author{Li Qiwei}
\affiliation{%
  \institution{University of Michigan}
  \city{Ann Arbor}
  \state{Michigan}
  \country{USA}}
\email{rrll@umich.com}

\author{Katelyn Kennon}
\affiliation{%
  \institution{University of Michigan}
  \city{Ann Arbor}
  \state{Michigan}
  \country{USA}}
\email{kkennon@umich.com}

\author{Nicole Bedera}
\affiliation{%
  \institution{Beyond Compliance}
  \city{Minneapolis}
  \state{Minnesota}
  \country{USA}}
\email{nbedera@umich.edu}

\author{Asia A. Eaton}
\affiliation{%
  \institution{Florida International University}
  \city{Miami}
  \state{Florida}
  \country{USA}}
\email{aeaton@fiu.edu}

\author{Eric Gilbert}
\affiliation{%
  \institution{University of Michigan}
  \city{Ann Arbor}
  \state{Michigan}
  \country{USA}}
\email{eegg@umich.edu}

\author{Sarita Schoenebeck}
\affiliation{%
  \institution{University of Michigan}
  \city{Ann Arbor}
  \state{Michigan}
  \country{USA}}
\email{yardi@umich.edu}

\renewcommand{\shortauthors}{Qiwei et al.}

\begin{abstract}

Non-consensual intimate imagery (NCII), also known as image-based sexual abuse (IBSA), is mediated through online platforms. Victim-survivors must turn to platforms to collect evidence and request content removal. Platforms act as the crime scene, judge, and jury, determining whether perpetrators face consequences and if harmful material is removed. We present a study of NCII victim-survivors' online reporting experiences, drawing on trauma-informed interviews with 13 participants. We find that platform reporting processes are hostile, opaque, and ineffective, often forcing complex harms into narrow interfaces, responding inconsistently, and failing to result in meaningful action. Leveraging institutional betrayal theory, we show how platforms' structures and practices compound harm, and, in doing so, surface concrete intervention points for redesigning reporting systems and shaping policy to better support victim-survivors.
\end{abstract}

\ccsdesc[500]{Human-centered computing~Empirical studies in collaborative and social computing}
\ccsdesc[500]{Human-centered computing~Collaborative and social computing theory, concepts and paradigms}
\ccsdesc[500]{Human-centered computing~Social media}
\ccsdesc[300]{Human-centered computing~Empirical studies in HCI}
\ccsdesc[500]{Human-centered computing~Social networking sites}

\keywords{Social media, Social Computing, online abuse, online sexual violence, institutional betrayal, NCII, IBSA, NCIM}

\received{11 September 2025}

\maketitle

\section{Introduction}

As social lives unfold online, non-consensual intimate imagery (NCII) has become a pervasive form of abuse. NCII refers to non-consensual creation, distribution, or threats to distribute sexual content, including sextortion, recordings of sexual assault, sexual deepfakes, and what is commonly known as ``revenge pornography''~\cite{mcglynn_beyond_2017}. Like contact sexual assault, NCII is a violation of consent and bodily autonomy. However, its digital manifestation enables abuse to persist long after the primary perpetrator’s departure, as content can be copied, re-uploaded, and amplified~\cite{qiwei2024sociotechnical,huber_non-consensual_2024}. NCII victim-survivors also endure compounded harms by multiple perpetrators such as networked harassment and doxing~\cite{mcglynn_its_2021}. Because NCII weaponizes the body itself, platform policies and systems designed to combat general forms of online harassment are inadequate to address this abuse~\cite{qiwei2024reporting}. Perpetration, harassment, and abusive posts associated with NCII all unfold on social media platforms. These platforms thus wield unprecedented power as crime scene, evidence locker, judge, and jury, simultaneously hosting the abuse, controlling access to evidence, and determining if and when harmful content is removed~\cite{gillespie_content_2020,gillespie2017platforms,qiwei2024reporting}. Their decisions shape whether victim-survivors experience resolution or face compounding harms. Because platforms hold this degree of power over victim-survivors’ recovery and safety, understanding their reporting processes is critical. Yet, despite NCII’s prevalence and legal recognition, little research has examined how victim-survivors actually experience reporting on online platforms. \textcolor{black}{Even less work has explored the additional harm that occurs when the very online platforms victim-survivors must rely on to address NCII fail to exercise their power to prevent or respond to abuse or even exercise it in ways that worsen victim-survivors' outcomes. This phenomenon has been examined in traditional institutional settings as institutional betrayal, but has yet to be extended to platforms as institutions.}

This paper answers the following research questions: 

\begin{itemize}
    \item[]\textbf{RQ 1:} How do victim-survivors navigate platform reporting systems for NCII?
    \item[]\textbf{RQ 2:} In what ways do platform reporting processes constitute institutional betrayal?
\end{itemize}

Drawing from trauma-informed interviews with 13 participants, we present a study of NCII victim-survivors’ reporting experiences with online platforms through the lens of institutional betrayal, a psychological theory that examines how institutions (re)traumatize those who belong to them in the aftermath of traumatic events~\cite{freyd_betrayal_1996,smith_dangerous_2013,bedera2022illusion,smith2017insult}. By applying this framework, we illustrate that platform reporting failures are not merely usability flaws or design oversights. Rather, platforms, as powerful digital institutions, compound the primary harm of NCII by imposing additional trauma via their reporting tools and processes, reproducing and extending violence against victim-survivors.

We find that victim-survivors report online to remove abusive content and to stop perpetrators from causing further harm. Platforms create hostile, opaque, and ineffective processes. Reporting requires re-engagement with abusive content, interfaces force victim-survivors to compress their experiences to fit narrow categories, and requests are often met with denial of legitimacy, abandonment, or even punitive action. Instead of offering redress, these responses ultimately drive many victim-survivors off the platforms they previously depended on for safety and social connection.

Identifying \ib in platforms allows us to render the NCII harms caused by platforms as distinct from those caused by perpetrators. This distinction reveals that a portion of harm experienced by NCII victim-survivors is produced by platforms acting as institutions and, therefore, may be amenable through design and policy choices. \textcolor{black}{In addition, our research protocol, designed in collaboration with trauma experts, constitutes both an ethical and methodological contribution to the literature in its prioritization by design of participant autonomy, choice, and safety~\cite{alessi2025applying,chen2022trauma}. Given that NCII victim-survivors previously encountered trauma and betrayal in online environments, and that recruitment and interviewing often occur in similar venues, we offer our protocols as a model for future researchers undertaking similar work with this population, in the hope that such approaches can help foster online research encounters that are, by contrast, reparative and empowering.}

In summary, we contribute 1) empirical insights into how victim-survivors report NCII to online platforms, 2) theoretical insights into the application of institutional betrayal theory to online platforms as institutions, 3) a trauma-informed methodology designed to mitigate (re)traumatization of multiply-betrayed NCII victim-survivors and researchers, and 4) design and policy directions that follow from the application of an \ib framework.

In what follows, Section \ref{rr} provides background on the harms experienced by NCII victim-survivors, the challenges of reporting online, and the foundations of institutional betrayal theory. Section \ref{methods} details our methods, analytic approach, and ethical considerations. Section \ref{results} presents empirical and theoretical findings that emerged from our data, as well as changes that some victim-survivors wished to see implemented in reporting processes. Finally, Section \ref{discussion} examines \ib in platforms and outlines its implications for both design and policy.

\begin{figure}
    \centering
    \includegraphics[trim={400 0 1650 0}, clip, width=0.7\linewidth]{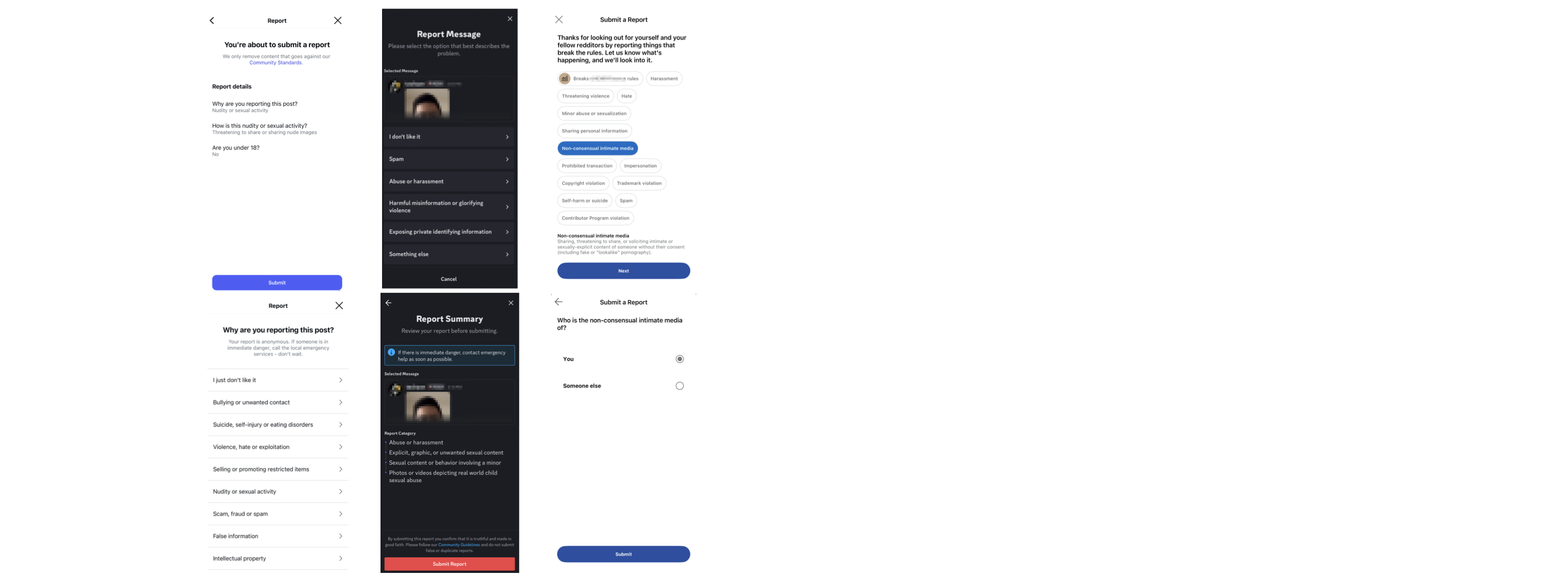}
    \caption{Process for making a report on a post or message on Instagram (left), Discord (middle), Reddit (right).}
    \label{fig:reportingUI}
\end{figure}

\section{Related Research} \label{rr}

We review the distinct nature of NCII harms, how victim-survivors navigate fragmented help-seeking and reporting pathways, and how \ib theory explains the (re)traumatization that often follows.

\subsection{Non-consensual intimate imagery harms and help-seeking} 

Non-consensual intimate imagery (NCII), including so-called ``revenge pornography,'' sextortion, covert recordings, and AI-generated deepfakes, represents a growing and rapidly evolving form of online sexual violence. NCII typically involves the creation, distribution, or threat of distributing intimate images, videos, or other forms of media without the consent of the person(s) depicted. Although its digital nature introduces unique affordances for harm, NCII shares key similarities with physical sexual assault (e.g., unwanted touching, rape) in terms of its causes, prevalence, and impacts~\cite{mcglynn_beyond_2017}. NCII can also co-occur with physical sexual violence, as in cases where a sexual assault is filmed. Research highlights the widespread prevalence of NCII, finding that more than 1 in 5 adults reported at least one experience of victimization~\cite{umbach_prevalence_2025}. Victimization is disproportionately concentrated among women, LGBTQ+ individuals, younger adults, disabled people, and Indigenous populations~\cite{flynn_intersectional_2024, umbach_prevalence_2025, henry_image-based_2019, ray_sextortion_2025}. NCII victim-survivors experience multi-faceted harms comparable to, or exceeding, those associated with physical sexual assault, leading to a range of adverse mental health outcomes, including depression, anxiety, post-traumatic stress disorder (PTSD), suicidality, self-harm, and disordered eating~\cite{campbell_social_2022,ditullio_feminist-informed_2019, huber_shadow_2023, ray_sextortion_2025, siegel_anything_2024, snaychuk_technology-facilitated_2020}. NCII victim-survivors often retreat from both online and offline spaces, leading to what has been described as a ``social rupture'' that disrupts bodily autonomy, identity, and interpersonal connection~\cite{mcglynn_its_2021}. 

The digital nature of NCII introduces new dimensions of harm. While physical sexual violence occurs at a specific time and place, the technological infrastructure of social media and internet platforms fundamentally transforms how, when, and where online sexual violence manifests. Content can be instantly duplicated across multiple platforms and shared through anonymous networks, victim-survivors may be identified and doxxed, and harm is amplified by platform algorithms that promote engaging content regardless of its harmful nature~\cite{qiwei2024sociotechnical,citron2014hate,citron2022fight}

Yet, public discourse reinforces the framing of NCII as less serious than physical sexual assault, discouraging many victim-survivors from seeking even informal support~\cite{harper_development_2023,fahmy_comparative_2024, mcglynn_beyond_2017}. Even when victim-survivors believe that their experiences warrant reporting, they often lack clear information about how to proceed or feel skeptical about available options \cite{colburn_help-seeking_2023, flynn_deepfakes_2022, campbell_social_2022}. Although civil remedies may offer more flexibility than criminal processes, they are often unaffordable and inaccessible, with fewer tailored legal or financial supports available for NCII than to address harms such as intimate partner violence~\cite{eaton_victim-survivors_2024, rackley_seeking_2021}. When victim-survivors do engage with criminal legal systems, their experiences often confirm their fears, as legal tools rarely translate into accessible, survivor-centered reporting pathways~\cite{qiwei2025law}. In one survey, fewer than 30\% of adults who reported NCII to police found it helpful, citing minimization, victim blaming, and inaction~\cite{colburn_help-seeking_2023}. While reporting, victim-survivors may be required to repeatedly relive their victimization and share intimate images with police, an experience many find deeply distressing~\cite{eaton_victim-survivors_2024}. 

Crucially, most of the NCII help-seeking literature comes from psychology and law, with limited contributions from HCI. This gap is striking given that NCII unfolds in online spaces. HCI has not examined how platforms themselves can shape or suppress help-seeking, with existing work on online harms addressing only peer-to-peer support~\cite{wei_understanding_nodate}. What remains underexplored is the role of platforms as mediators, how people engage with them when seeking help, and how platform processes enable or block pathways to redress, creating new harms for victim-survivors. 

\subsection{Reporting sexual violence online}

Although NCII is criminalized in the U.S., legal options overwhelmingly target perpetrators rather than guaranteeing the removal of harmful content, a task rendered complex by the legal landscape online. There is no mechanism that ensures distributed NCII duplicates are located and removed on behalf of victim-survivors when perpetrators face legal consequences ~\cite{qiwei2025law}. In May 2025, the TAKE IT DOWN (Tools to Address Known Exploitation by Immobilizing Technological Deepfakes on Websites and Networks) Act became U.S. law. \textcolor{black}{The Act requires online platforms to, by May 2026, remove NCII no more than 48 hours after receiving a report from a victim-survivor or an authorized agent. It also obligates platforms to make reasonable efforts to locate and remove any known identical copies of the reported material, though it does not specify further what ``reasonable'' entails or how it is to be measured. Enforcement falls under the Federal Trade Commission (FTC), which may bring civil actions against platforms for noncompliance, including fines and injunctions~\cite{take_it_down_2024}.}

While the Act is promising at face value, questions remain about who benefits, how enforcement will unfold, and whether platforms will meaningfully comply~\cite{TAKEITDOWN_Act_2025,grimmelmann2025deconstructing}. A longstanding barrier to content removal and platform accountability is Section 230 of the U.S. Communications Decency Act, which shields platforms from civil liability for user-generated content, positioning them as neutral ``messengers'' rather than accountable custodians~\cite{gillespie2018custodians,gillespie2017platforms}. How the TAKE IT DOWN Act will intersect within Section 230 remains uncertain, particularly given that it proposes civil enforcement mechanisms. As a result, victim-survivors still cannot rely on platforms to act. One notable workaround is the Digital Millennium Copyright Act (DMCA), which allows some victim-survivors to assert copyright ownership over their content to directly request removal. Despite its limitations, including accessibility hurdles and risks of privacy exposure, the DMCA remains the primary pathway for removal recommended by legal experts~\cite{qiwei2025law,qiwei2024reporting,ccri_online_removal}. 

\textcolor{black}{Section 230 was previously narrowed by SESTA-FOSTA (the Stop Enabling Sex Traffickers Act and the Fight Online Sex Trafficking Act), a pair of 2018 laws representing the first major rollback of Section 230’s protections. SESTA-FOSTA was ostensibly intended to combat the use of online platforms for sex trafficking by increasing platforms’ civil and criminal liability for knowingly assisting, supporting, or facilitating trafficking or financially profiting from it. However, the laws have been widely criticized for their overly broad definitions of both ``trafficking'' and ``facilitation,'' prompting many platforms to ban or heavily restrict content related even to consensual sex work or legal sexual expression~\cite{blunt_erased_2020,are2022shadowban}. Critics argue that the laws produced a chilling effect on sexual speech online and may have, in fact, increased risks for both trafficking victims and consensual sex workers, including those victimized by NCII~\cite{kinzer_policy_2025, barwulor_disadvantaged_2021}.}

Despite HCI’s extensive engagement with content moderation, platform governance, and online harassment, its literature has not addressed victim-survivors' experiences reporting NCII. Prior research has examined online harassment broadly, including online tools for reporting street harassment, such as Hollaback!, and victimization of vulnerable groups such as female journalists~\cite{dimond2013hollaback, goyal_you_2022,han2024pressprotect}. Studies underscore challenges such as the difficulty of documenting harassment, the stress of reporting under duress, and opaque or inconsistent reporting processes~\cite{goyal_you_2022,matias_preventing_2019, matias_reporting_2015}. Perpetrators’ use of networked harassment and anonymity exacerbates these problems, while the burden of evidence collection often falls on victims themselves~\cite{nova_online_2019}. While prior work on online harassment provides critical foundations, it typically does not address the distinct dynamics and stakes of NCII. The unique role of intimate content introduces additional layers of harm and complexity that require dedicated study. Compared with other forms of online harassment, NCII is more persistent and invokes more urgency, as abusive media are a violation of both privacy and bodily autonomy that can be copied and disseminated indefinitely, which results in takedown processes for NCII being highly burdensome~\cite{huber_non-consensual_2024}. Work in HCI on NCII reporting is sparse. De Angeli et. al's work is limited to observing UI in controlled settings, while Qiwei et al. provide data on real takedown timelines but do not examine the reporting experience itself~\cite{qiwei2025law,qiwei2024reporting,de_angeli_reporting_2023}. Together, these gaps highlight the need for empirically grounded, survivor-centered research on NCII reporting in HCI.

\subsection{\IB}

Institutional betrayal occurs when an institution's actions---or inactions---in response to a traumatic event exacerbate a victim-survivor's original trauma. ~\cite{smith_institutional_2014,freyd_betrayal_1996,smith_dangerous_2013}. After sexual harm, victim-survivors often turn to institutions such as schools, workplaces, or churches to seek support, protection, and formal accountability for perpetrators. Institutions have immense power, giving them the ability to upend power disparities that allow sexual harm to occur or persist. Further, institutional interventions are often necessary for a victim-survivor to seek redress that they lack the power to access as individuals. Rather than receiving the care they expect, many victim-survivors instead experience an additive layer of trauma when institutions dismiss, blame, neglect, or even attack them. These harms are systemic, embedded in institutional norms and practices. In essence, the institution becomes another perpetrator with the potential to cause trauma akin to a ``second rape''~\cite{campbell_preventing_2001}.

Institutional betrayal research has focused on how organizations produce and exacerbate broad types of interpersonal harm (sexual violence, racial discrimination, environmental hazards, ableism) across a wide range of traditional institutional settings (schools, military, churches, workplaces, healthcare systems)~\cite{christl_when_2024,smidt_out_2021,gardner2022institutional,holliday2019seeking}. Institutional betrayal in these contexts has most commonly been measured using a version of the Institutional Betrayal Questionnaire (IBQ), a self-report instrument consisting of items that describe possible institutional behaviors related to traumatic or harmful events~\cite{smith_dangerous_2013,smith2017insult}. The IBQ asks, for example, whether an institution ``[created] an environment in which this type of experience seemed common or normal,'' ``[denied] the experience in some way,'' or ``[punished respondents] in some way for reporting the experience.''~\cite{smith2017insult} Higher total scores indicate greater perceived institutional betrayal, but endorsement of any single item constitutes evidence of betrayal~\cite{smith2017insult}. \textcolor{black}{In addition to these core items, respondents are asked whether the institution was one they ``strongly identified with or felt part of'' and whether they remain connected to it. These questions are designed to assess respondents' trust and dependence on the institution, essential preconditions for later betrayal. Across studies, the IBQ has demonstrated good construct validity and internal consistency~\cite{smith_dangerous_2013, smith_institutional_2014, smidt2021out, gardner2022institutional}}

\textcolor{black}{The IBQ has since been adapted for use in diverse institutional settings and populations. The updated IBQ-2 is a broader and psychometrically strengthened version that expands the scope of measured betrayal behaviors to better include passive forms of harm and systematic and environmental contributors (i.e., exclusion, devaluation), in addition to more overt institutional actions (i.e., retaliation, cover-ups) ~\cite{smith2017insult, wolff2025ibq2m, boyd2023psychiatric}. IBQ-2 consists of 12 standardized items, rated on a Likert scale rather than dichotomously, providing a more nuanced assessment of perceived betrayal. At least one psychometric evaluation of IBQ-2 supports a two-factor structure that distinguishes between the institutional promotion of victimization (i.e., creating and enabling conditions that produce victimization) and institutional responses to victimization (i.e., reactions---or inaction---after harm has occurred)~\cite{reffi_psychometric_2021}. Despite this early evidence, researchers generally recommend the use of the full IBQ-2 rather than separate subscales~\cite{kelley2018mst}.}

Specifically in the context of sexual violence, institutional betrayal experiences are both common and devastating. Some studies document rates of 45 to 60\%, while others report that nearly 100\% of their sample experienced institutional betrayal~\cite{christl_when_2024, lorenz_title_2021}. Institutional betrayal in these cases compounds the original violence because it echoes its dynamics; the betrayal is perpetrated by a trusted and powerful entity that victim-survivors expected would care for them. ~\cite{freyd_betrayal_1996}. This entanglement worsens mental and physical health outcomes, increases self-blame, and reduces help-seeking, beyond the impacts resulting from the initial trauma~\cite{smith_dangerous_2013}. Institutions also possess the power to induce tangible losses for victim-survivors, including school drop-out, job loss, ostracization, and lost advancement opportunities~\cite{christl_when_2024, smith_institutional_2014}. Notably, queer survivors and people of color are at increased risk of both institutional betrayal and its psychological toll, perhaps owing to their already-marginalized position within institutions~\cite{smidt_out_2021, smith_sexual_2016, gomez_microaggressions_2015}.

Despite parallels between contact sexual violence and NCII, online platforms have not been examined as potential perpetrators of institutional betrayal. However, like traditional institutions, online platforms function as gatekeepers of justice, social support, and legitimacy, representing a ubiquitous presence in the lives of victim-survivors. As such, this study explores whether platforms enact similar betrayals toward those who depend on them.

\section{Methods} \label{methods}

This study investigates how victim-survivors pursue the removal of NCII and obtain evidence from online platforms. We conducted semi-structured interviews (n=13) with adult participants residing in the U.S. between April and August 2025. The research was determined to be exempt by our institution's IRB. 

\subsection{Recruitment}
We recruited participants who had experienced the non-consensual creation or sharing of intimate photos or videos of themselves online. Eligibility was limited to adults (18 years or older), although many participants reported incidents that occurred during their adolescence. We focused on U.S.-based participants because legal frameworks governing both NCII perpetration and the treatment of online content vary significantly between countries. Recruitment posts and advertisements read ``Have your intimate videos or photos been shared online without consent?'' and avoided using loaded words such as ``survivor'' so that we did not impede self-labeling. We recruited through multiple channels:

\begin{itemize}
    \item[] \textit{Social media advertising:} Deployed broad-targeted Instagram ads across the U.S. directed at men and women. 
    \item[] \textit{University research listservs:} Distributed through academic research participant pools to reach individuals experienced with study participation, but not necessarily with sexual violence. 
    \item[] \textit{Victim-survivor–focused social media:} Recruited through accounts with large followings among victim-survivor communities, particularly those engaged in conversations about sexual abuse \textcolor{black}{on X and Bluesky}. 
    \item[] \textit{Advocacy network:} Partnered with an advocacy organization that specifically supports NCII victim-survivors. This allowed us to reach individuals likely to be more experienced with reporting mechanisms and deeply impacted by their NCII victimization. 
\end{itemize}

\color{black} 
We selected a diverse group of participants---across gender, sexual orientation, race, and age at time of victimization---who had encountered diverse types of NCII harm (see Table \ref{tab:reports}). This enabled us to capture varied reporting experiences and to understand how these differences might shape reporting perceptions and outcomes. We aimed to recruit a large proportion of participants who had interacted extensively with multiple online infrastructures, including navigating different platform policies and requesting content removals, so that we could compare reporting experiences across platforms in greater detail. Simultaneously, in the service of breadth, we intentionally included participants with limited reporting experience, such as those who had reported to only one platform or submitted only one request to platforms. 

Institutional betrayal---particularly when it obstructs or discourages reporting---results in many victim-survivors disengaging from the reporting process before ever submitting a report or struggling to locate the avenue through which to do so. Focusing on individuals who at least partially engaged with a reporting process is therefore an important limitation of our recruitment strategy. Although identifying early pain points is essential for improving reporting systems, our study concentrated on the end-to-end reporting process to surface more points of friction. Notably, many participants had multiple discrete experiences of NCII victimization or were drawn back into the reporting process due to re-uploads. As a result, several could reflect on moments when they chose not to report their NCII again after previous negative experiences. These accounts offered partial insight into the early betrayals that can deter victim-survivors from reporting at all.

\color{black}
We initially reached out to potential participants via email or text message according to their stated preference. Potential participants were first offered an optional 15-minute pre-interview conversation, allowing them to meet interviewers, ask questions, and ensure comfort before choosing whether to proceed with the full interview. Participants chose their preferred interview modality over Zoom, either audio-only or video. All interviews were conducted in English with two interviewers in the Zoom room: one researcher and one trauma-informed research consultant to reduce the risk of re-traumatization during data collection and provide immediate support in the case of traumatic reactions. Interviews lasted an average of 106 minutes and, with the exception of one interview, lasted between 90 and 240 minutes. Participants received compensation of \$100 in the form of a Visa gift card. Following each interview, audio recordings were transcribed via Zoom then manually cleaned and reviewed by research assistants trained to transcribe data on sensitive topics. 

\subsection{Analysis}
\color{black}
Our analysis combined deductive and inductive approaches. {The Institutional Betrayal Questionnaire-2 (IBQ-2), a validated 12-item self-report measure designed to capture both passive and active forms of institutional betrayal, served as the deductive coding scaffold~\cite{smith2017insult}. While participants did not complete the IBQ-2 themselves, its items informed the initial codebook and guided our identification of institutional betrayal within the qualitative data. Two authors independently coded three randomly selected interview transcripts using the IBQ-2 betrayal behavior items as deductive codes to identify evidence of institutional betrayal. For example, we coded instances where online platforms failed to ``take proactive steps'' to prevent NCII, ``denied experiences'' of NCII, or ``created an environment where continued membership was difficult'' for participants. Because institutional betrayal presupposes a relationship of trust in or identification with an institution---as reflected in IBQ-2's supplemental items---we also coded for moments when participants described their relationship to platforms, including expressions of dependence or perspectives on the platforms' roles in their lives. Coding discrepancies were reconciled through discussion and comparison, consistent with best practices in thematic analysis~\cite{terry2017thematic,braun2006using}.

Following agreement on the coding framework, all transcripts were reviewed for content corresponding to IBQ-2 items. Nearly all IBQ-2 items were reflected in participants' accounts, particularly item 4 (making it difficult to report) and item 5 (responding inadequately to reports). We then conducted inductive thematic analysis within each IBQ-2 aligned category to identify additional, platform-specific patterns of institutional betrayal~\cite{fereday2006demonstrating}. Two authors then applied \ib to affordances and governance structures of digital systems (See Table \ref{tab:betrayal}). This hybrid deductive-inductive approach allowed us to assess how established dimensions of institutional betrayal theory apply to platform settings, while also allowing unique expressions of these harms to emerge, specific to online institutions. 

\color{black}
\subsection{Ethical Considerations} 

An ethical approach to research is not about ``not doing harm,'' it is about restoring agency to a group of people who have had it revoked ~\cite{tronto1993political,ellis2007telling}. The IRB similarly acknowledges that, while research inherently involve risks, they must be reasonable in relation to the anticipated benefits~\cite{hhs_common_rule_46_111}. Our participants represent a highly vulnerable and stigmatized population. We prioritized their confidentiality, emotional safety, and autonomy at every stage of the study. To safeguard participant identities and protect them from further harm, we implemented strict privacy protocols. Participants were encouraged to use pseudonyms and alternate email addresses for communication to maintain anonymity and avoid re-triggering past trauma. We sought input from advocacy professionals and individuals with lived experience to review our study materials for ethical integrity and risk sensitivity. 

All interviews were conducted by the first or second author in conjunction with a trauma-informed research consultant with a crisis counselor certification, ensuring immediate support if participants experienced emotional distress. Participants needed more control over their participation than in typical research settings. We considered this need in consent forms, rights explanations, and data collection procedures. We designed outreach communications to clearly identify us as researchers and to avoid triggering language, acknowledging previous incidents where participants had experienced deceptive or harmful contact. We prioritized participants' choices and emphasized their role in shaping conversations. Participants were also provided with information about online safety resources and avenues for follow-up support, regardless of whether they completed the study. \textcolor{black}{Finally, we shared a draft of this article with all participants and provided time for them to review their representation and provide feedback. All participants who responded expressed positive reactions to their portrayals; several noted that reading about others' similar experiences had been healing in that it reduced their sense of isolation.} This aligns with literature suggesting that research participation promotes healing for trauma survivors because it offers opportunities to feel heard and empowered, particularly for sexual victimization that is perceived as taboo~\cite{newman2004risks,campbell2009rape}.

Working with traumatic narratives can lead to emotional fatigue, secondary trauma, and desensitization, any of which could affect how researchers interact with participants and interpret their stories~\cite{van2019secondary}. To mitigate these risks, we incorporated multiple safeguards into our study. First, all research team members received training in trauma-informed interviewing before developing study materials and engaging with participants. This training emphasized emotional attunement, non-extractive questioning, and maintenance of researcher boundaries. Second, we deliberately paced data collection to avoid burnout. Each researcher conducted no more than one interview per day and no more than three interviews per week. This structure allowed time for emotional processing and reduced the cumulative toll of repeated exposure to distressing content. Third, we held debriefing sessions after the conclusion of each interview, drawing on models from clinical and emergency response contexts~\cite{fullerton2000debriefing}. These debriefs offered space to reflect on emotionally difficult moments, share coping strategies, and make adjustments to our practices as needed. Finally, we extended these principles to transcription work. We engaged research assistants trained in social work and trauma-informed care for transcription. This ensured that all research tasks were aligned with our ethical commitments.

\section{Results} \label{results}

\reports

Participants described approaching reporting systems with different goals. Sometimes, they prioritized removal of abusive content. Other times, participants reported to document incidents or to stop continuous perpetration. These different orientations toward reporting shaped how participants navigated platforms and experienced their responses. We present our findings in four parts: Section \ref{barriers} on the difficulties of accessing and using reporting tools in the first place, Section \ref{responses} detailing the inadequate and often harmful responses returned by platforms, Section \ref{mistreatment} on the ways these processes left victim-survivors feeling mistreated and ultimately driving them away, and Section \ref{conditions} demonstrating how initial trust in and dependence on platforms as sites of redress gave way to betrayal after victim-survivors experienced reporting. \textcolor{black}{We recognize that an implicit component of this paper is educating readers about what the reporting process entails. Accordingly, we have organized our findings around the trajectory of this process for clarity.}

\bexamples

\subsection{Difficult reporting processes} \label{barriers}

Victim-survivors shared that it was difficult to report NCII to platforms, \textcolor{black}{consistent with \ib (see IBQ-2 Item 4 in Table \ref{tab:betrayal}). Initial encounters with reporting processes often left participants feeling disoriented, confused, and powerless because they struggled to find reporting options appropriate for their cases, felt unsafe while reporting, and were required to repeatedly relearn new reporting processes.} We describe the specific obstacles that participants faced in their first steps toward seeking help.

\subsubsection*{Forcing complex violence into a drop-down list}

Online abuse rarely maps cleanly into a singular category of harm, least of all from the perspective of the person living through it. Participants struggled to translate the violence they experienced into the narrowly defined categories offered by reporting interfaces. To make their experiences legible, victim-survivors engaged in trial and error, trying different strategies, pivoting when one path led to a dead end, and reframing their experiences to appear ``legitimate'' to platforms, echoing observations by Felstiner et. al on the difficulties encountered in ``naming, blaming, and claiming'' in other institutions' reporting processes~\cite{felstiner2017emergence}. 

Paradoxically, participants believed reports framed as a singular violation were taken more seriously than those that captured the full range of overlapping harms. This forced victim-survivors into a painful choice: distort and minimize their experience to be heard or risk dismissal~\cite{crenshaw2013mapping}. An anonymous perpetrator stole P11’s erotic photos, created an impersonation account on Instagram, and hosted a private website to solicit sex from P11’s followers. This case simultaneously involved impersonation, harassment, and sexual content violations, yet P11 could not select all these categories at once when reporting. In fact, they believed they risked failure if they labeled the incident differently across reports. See Figure \ref{fig:reportingUI} for examples of reporting interfaces. 

\begin{quote}
    You can report for impersonating a person, you can report for harassment, you can report for inappropriate content, and I remember searching online to figure out which of those is the most effective, because it felt like there were multiple violations happening here, and I wasn't sure which was most likely to get listened to.
\end{quote}

In other cases, the harm was not in the posts themselves but in the context of their circulation. For P1, photos safe to share in some settings became deeply unsafe when reposted by an impersonation account targeting the Muslim community. The enforcement of narrow reporting categories reduces harm to isolated accounts or images, not accounting for the broader social and cultural contexts that make these violations dangerous~\cite{shahid2023decolonizing}. 

\begin{quote}
    I live like a double life. If the audience for that was not [NATIONALITY] and Muslim people [the violation would not be] to the same extent \ldots the pfp was a picture of me in leggings, and you could see my butt. I don't like the idea of [NATIONALITY] people seeing that. I'm okay with, people that are used to seeing women dressed in more revealing stuff seeing me like that. So that just felt a lot more like shameful, shorts, tank tops, bodycon dresses. 
\end{quote}

\textcolor{black}{Participants' experiences often defied or exceeded the narrow, fixed categories that platforms offered. This forced them into unnecessary choices that made reporting more difficult and raised doubt about whether platforms would take their reports seriously. In reducing multidimensional violations to rigid, one-dimensional items, platforms effectively denied victim-survivors yet another form of autonomy as they attempted to seek help.}

\subsubsection*{Design flaws undermined safety}

Even the most basic design flaws---ones long solved in other products---made reporting NCII difficult. Participants described being constrained by character limits that reduced their violation to two sentences or finding that reporting interfaces simply did not function on mobile. This was especially significant given that many reports needed to be filed quickly to prevent further dissemination.

More troubling were cases where self-protective actions taken by victim-survivors inadvertently made reporting impossible. P4, whose perpetrator threatened to share her nude photos on Reddit and Discord, described blocking and unblocking him multiple times as a safety tactic. However, on Reddit, this back-and-forth prevented her from filing a report against him. Similarly, on Discord, deleting hard-to-revisit abusive messages also erased the evidence needed to submit a report. 
\begin{quote}
    From what Discord shows, you can't see deleted messages and images \ldots I wish that Discord would change their DM reporting system \ldots [so] I could report a full DM even including deleted messages. 
\end{quote}

\textcolor{black}{Protective steps were sometimes directly at odds with what the platform required to submit a report. In these moments, platforms forced victim-survivors into an impossible trade-off: protect themselves now or preserve evidence for a system that might not help them.}

\subsubsection*{Reporting felt unsafe}

When reporting was possible, participants described approaching the process with extraordinary caution because of the sensitive nature of the materials involved. This included links to abusive intimate content, screenshots of their nude photos, and written descriptions of the abuse. P10's intimate photos were non-consensually shared with Discord group members, one of whom later messaged him the image on Instagram as a threat to disseminate. When P10 attempted to report on Instagram and Discord, he felt compelled to move with extreme precision and care. Any misstep---a wrong click, a premature submission---risked either sabotaging content removal or even alerting the perpetrator to his engagement with the content.

\begin{quote}
    I had to be precise and methodological about reporting the picture \ldots Sometimes you get water on your phone, or you swipe wrong, or you accidentally misstep, and suddenly send the message. If I had made a misstep, misclicked, liked it, or shared it or something, then the other guy would probably get a notification about it and that would land me into more trouble. 

\end{quote}

Reporting required victim-survivors to directly re-engage with abusive materials, a process that often felt like reproducing the very harm they were trying to stop. To request takedowns, participants had to submit their own images---sometimes searching through dozens of pages of other victims' NCII to locate them. This forced re-exposure caused significant distress. Participants described the irony of having to re-circulate content that was never meant for others' eyes. P8's nude photos were stolen during a Snapchat breach, and their images were posted on NCII-hosting sites alongside identifying information. Because P8 no longer had access to the photos on their device, they could only report the content by first locating and screenshotting it on the host website. P8 quickly deleted the images again after reporting, but went through the process five more times when reporting failed or content was re-uploaded.

\begin{quote}
    I have to screenshot [the photos] and then submit them. How do I know that Google is not gonna get hacked, and then another person's gonna take the images and do something with them? \ldots It was triggering. I'm going through my phone. I have to look at the pictures that I don't want to look at. It was traumatizing to look at them again.
\end{quote}

\textcolor{black}{These fears---that a perpetrator might be alerted when a report is filed, and that reporting requires repeated contact with the very materials they want gone---made the act of reporting itself deeply distressing. In the moments they were trying to protect themselves, participants were instead exposed to new risks and retraumatization.}

\subsubsection*{Reporting as collective labor}

Reporting takes a village, often requiring both persistence and repetition. Participants described having to submit the same report multiple times before platforms took action. In some cases, they enlisted friends to help amplify their efforts, asking them to file duplicate reports in hopes of forcing a response. This collective strategy was only possible when participants trusted their social circle not to judge them or when the violation involved impersonation accounts that did not display nude images publicly. Most of the time, however, victim-survivors lacked the ability to mobilize such support and bore the reporting burden alone. P11, who was impersonated, began receiving messages from friends warning them about the fake account. In response, they rallied friends to report the impersonation.

\begin{quote}
    Everyone who messaged me [about the impersonation account] I was like, please report it \ldots Then I messaged another 10, 15 of my friends and asked them to go report this account. And then myself, obviously, and I did it every day for like \ldots until it came down.
\end{quote}

\textcolor{black}{Participants saw collective reporting as one of the few ways to make platforms pay attention and increased the chances they would take action. But relying on others in this way invoked its own cost by requiring participants' to undertake the labor of both crisis coordination and risk management.}

\subsubsection*{Inconsistency across websites}

For participants whose media spread across the Internet, the reporting burden intensified quickly. They were forced to file separate reports on multiple websites, each with its own rules and requirements. Because every website had a different interface, victim-survivors had to learn and re-learn procedures under conditions of acute stress. P5, whose NCII circulated widely and was re-uploaded multiple times, described this exhausting cycle:

\begin{quote}
    The takedown request is different on every website \ldots once I've like mastered it for one website, then it would go up on another one, and then I'd be frantically trying to figure out their processes.
\end{quote}

\textcolor{black}{While earlier examples described participants submitting a large volume of reports on a single platform, NCII was often not contained to one site. Moving across platforms required participants to undertake the difficult labor of becoming lay experts on multiple platforms' processes, which they found mentally and emotionally exhausting.}

\subsection{Inadequate platform responses to reports} \label{responses}

Submitting a report was only the commencement of a challenging process. Even when participants went to distressing lengths to file, their reports were often mishandled or received an inadequate response, including no response at all \textcolor{black}{(See IBQ-2 Item 5 in Table \ref{tab:betrayal})}. Overwhelmingly, participants felt betrayed and hopeless in the face of inadequate platform responses. 

\subsubsection*{Reports met with silence or delay}

Participants' most common complaints after submitting reports were that they received no response or took so long to be addressed that redress became meaningless. Across platforms, participants described expending significant effort on reporting, only to be met with silence. Delayed responses frequently came too late to prevent cascading harm; the content had already been distributed and reposted widely. P5 emphasized the importance of timeliness, explaining that delays from Tumblr during the initial dissemination of her content allowed it to proliferate onto downstream websites. By the time Tumblr acted, the damage was already done.
\begin{quote}
    Fuck these fucking platforms. [The photo] had been up for 2 weeks. It had been shared and downloaded so many times by that point \ldots there was no coming back from that. At that point I was like, this is my life forever now. 
\end{quote}

\textcolor{black}{Participants stressed that removals only help if they happen quickly, before a post receives a lot of views or is uploaded elsewhere. Late action, though better than nothing, feels futile to victim-survivors and conveys that platforms do not understand the harm of NCII or take it seriously.}

Obtaining basic information from online platforms can also take painfully long. P3 had intimate photos posted across Reddit, 4chan, Telegram, and \textcolor{black}{websites intended to share NCII}, alongside identifying information that was sent directly to people in \textcolor{black}{her} life via Instagram. \textcolor{black}{She and her} lawyer repeatedly requested information from Instagram to identify the perpetrator, but encountered weeks of resistance.
\begin{quote}
    Instagram is a fucking nightmare of hell, where they will do anything in its power to prevent you from getting access to information that will help you. It took almost 7 weeks, [multiple] lawsuits in [different] states for my lawyer to be able to get back who was sending those messages.
\end{quote}

Notably, only a few participants in our study had access to legal support. When lawyers intervened, platforms were far more responsive, and participants described relief that the burden of constant reporting and follow-up had shifted from their shoulders, even though their lawyers encountered many of the same systemic barriers and could not always convince platforms to intervene. Lawyers acted as intermediaries between victim-survivors and platforms, partially buffering the experience of institutional betrayal. Yet, this discrepancy further entrenches inequity given most victim-survivors could not afford legal support.

\subsubsection*{Active harm during report review}

Participants took particular issue with platforms' practice of leaving content visible during the reporting review period. Victim-survivors described heightened anxiety knowing that NCII continued to accumulate views, downloads, and shares while platforms deliberated. The consequences were severe: keeping reported NCII active turned delay into a direct mechanism of harm, guaranteeing wider distribution and making eventual removal feel hopeless. By the time any response came for P5's reports, the material had already circulated far beyond the original site, seeding reposts across the internet.
\begin{quote}
    If you're saying this was non-consensually posted, it should be taken down immediately while they review it. It should not stay up while they fucking review it.
\end{quote}

\textcolor{black}{Overall, participants felt their needs were not central to platforms' design or decisions. Review processes that kept content visible during review seemed optimized for convenience and ease of posting while directly undermining the safety of victims.}

\subsubsection*{Opaque and automated responses}

Participants described platforms' inconsistent and often late responses as also being opaque and impersonal. After navigating an already taxing process, victim-survivors were met with nothing more than an automated message, with no channel for follow-up questions, clarifications, or appeals. This lack of communication left them feeling dismissed and powerless. P11 captured this frustration when describing Instagram’s handling of an impersonation report. The automated reply lacked any sense of acknowledgment or accountability. For them, the absence of a ``human touch'' made clear that Instagram as a platform was unconcerned with them. 

\begin{quote}
    There was no even pre-written message that Instagram sent me to apologize. [It made me think] social media is not a humane space, like it's not meant to treat people with compassion necessarily. It didn't feel like there was a human on the other side. 
\end{quote}

\textcolor{black}{After pushing themselves through a difficult reporting process and revisiting painful, deeply personal evidence, participants felt their effort wasn’t rewarded with anything in return. In addition to not meeting their original goals, the reporting process also made platforms and their experiences on them feel less human (see also section \ref{whatpeoplewant}) }

\subsubsection*{Insufficient actions}

When platforms did take action, participants often found their measures insufficient to address their concerns or prevent future harm. This insufficiency took two main forms. First, platforms removed the reported content but failed to impose consequences on infringing users. In these cases, perpetrators were free to create new accounts and continue posting, with no systemic safeguards to prevent recidivism. Second, when sanctions were applied, they were strikingly minimal. P11 finally received a response from Instagram about the impersonation account, only to learn that the account had been disabled for three days. They described this as a mere ``slap on the wrist'' for an offense that had caused lasting distress.
\begin{quote}
    It was just like hell. It was one of those micro actions that Instagram sometimes, if you'll comment something, the filters will auto remove it or they'll put a temporary hold on your account and say you have 3 days until you can post a story again. So they hadn't done any real action. 
\end{quote}

Participants pointed to similar gaps in external removals. P8 found it troubling that Google indexes sites that exist entirely to facilitate abuse at all, which grants them visibility and legitimacy.
\begin{quote}
     What do you mean that you can just have these websites on Google? I've always just thought that was weird and icky \ldots It's frustrating that Google allows these websites to even be a part of Google. 
\end{quote}

\textcolor{black}{Participants often felt that what platforms called ``action'' didn’t amount to real protection. Content came down, but the user posting it could continue. When platforms did impose penalties, they were so minor that they felt meaningless. Together, these responses left participants feeling dismissed and exposed to the same risks, all over again.}

\subsubsection*{Recurrent re-uploading} 

Because NCII can be copied and re-uploaded instantly, content removal often marked only a temporary reprieve before the beginning of the next round of harm. Victim-survivors described how new uploads would surface soon after platforms took action, forcing them to restart the entire process of locating the content, filing reports, and waiting for decisions, as if none of their prior effort had made a difference. This is illustrated by P8 as they describe what happens after a temporarily successful removal of a link from a hosting website through Google. 

\begin{quote}
    But then the websites would realize that [the photos] were taken down, and then they would just upload them again. So it was a never ending cycle \ldots I think I completed the revenge porn form 4 to 6 times. I was doing it probably every 48 to 72 hours of just constantly trying to get them down. Sometimes [a photo] would come down for 24 hours, and it'd be back. So then I'd have to do that one again. 
\end{quote}

\textcolor{black}{For many, every takedown simply portended more work in the future. Beyond a mishandling or inadequacy, this pattern, for some participants, began to more closely resemble a cover-up, in that the true content removal they pursued through the advertised process seemed impossible.}

\subsection{\textcolor{black}{Mistreatment by platforms}} \label{mistreatment}
Beyond these challenges, victim-survivors experienced institutional betrayal when their reports were denied, minimized, or met with punitive actions \textcolor{black}{(see IBQ-2 Items 5, 9, and 12 in Table \ref{tab:betrayal})}. Instead of feeling protected, victim-survivors felt blamed, silenced, or punished for seeking help---the same fears that often made them reluctant to seek help from other institutions in the first place.  

\subsubsection*{Denying the violation}

Just as traditional institutions often deny, dismiss, or obscure victim-survivors’ reports, platforms often refuse to recognize participants' experiences as violations, claiming that reported material ``does not violate terms of service.'' This denial was especially frustrating when the harm was clear to the victim-survivor, yet rendered invisible by the platform's seemingly arbitrary standards. P3 captured this frustration after reporting accounts that used \textcolor{black}{her} nude photos as a profile image and messaged \textcolor{black}{her} slurs like ``slut,'' along with other threats. For P3, the platform’s dismissal compounded the original violation by refusing to acknowledge the legitimacy of the harms \textcolor{black}{she} experienced.
\begin{quote}
    The fact that \ldots [Instagram] decided it doesn't violate the terms and conditions. Well then maybe the CEO should be revisiting those terms and conditions, because that's not right.
\end{quote}
\textcolor{black}{Denial was not experienced by participants as a mere misclassification. Instead, it sent a powerful institutional message that victim-survivors' NCII experiences did not matter or count as violations, signaling a withdrawal of support from a relied-upon platform.}

\subsubsection*{Punished for reporting}
At times, victim-survivors were punished for reporting abuse. After reporting a perpetrator on Discord who solicited CSAM from her in return for payment, P4 was banned by server moderators for ``unsafe activity'' and ``teen self-endangerment,'' leaving her locked out of the platform for several days. P4 felt that Discord treated her with more suspicion than they treated the perpetrator: 
\begin{quote}
Don't ban me first without telling me the situation has been taken care of. Like, we banned this guy's IP because it's like, I'm technically, you know, the victim here. You should also deal with the perpetrator.
\end{quote}
Discord's response also shaped how P4 approached reporting elsewhere. When later reporting to Reddit, she found herself censoring her description of events because she feared having her account there banned ``like Discord did" and did not want to experience the platform's ``extra judgment." At least partially due to Discord's reaction, P4 reported only fragments of what she experienced, worried that she would be held responsible for something criminal, despite the clear power imbalance and legal concerns in her situation. 

In other cases, punishment was broader and systemic. In response to SESTA–FOSTA, Tumblr removed all nude content rather than address the problem of non-consensual nudity ~\cite{blunt2020erased}. P5, whose NCII was originally posted on Tumblr, attempting to have the content removed from the site for years, even preparing a lawsuit against the platform. She ultimately abandoned the case after Tumblr's sweeping response to SESTA-FOSTA finally forced the content to be removed. Reflecting on the process, P5 noted that, even after all her investment to that point, the passage of SESTA-FOSTA made it ``too much work to try to get that case to go through."

For several participants, including P5, the ability to create and share consensual sexualized content via technology was an important part of reclaiming healthy sexuality, whether through nudes, burlesque, modeling, or pornography. P5 described sexting a partner again for the first time, years after her NCII ordeal began:
\begin{quote}
\ldots I was like, that's my marker. Like that's my point of  I have healed, like I have moved through this.
\end{quote}
The form of collective punishment enacted through SESTA-FOSTA, where the presence of perpetrators prompted blanket restrictions on all users, functioned as backlash against victim-survivors who sought safety and justice while also erasing those who shared consensual intimate images, including sex workers and others exercising bodily autonomy through image sharing. 

Platforms' retaliatory actions transformed help-seeking into an additional site of vulnerability for victim-survivors, reinforcing their sense that platforms are aligned with perpetrators rather than concerned with retaining them as community members. Even when participants did not experience direct retaliation from platforms, some described limiting or avoiding further reporting because they feared effective retaliation or counter-complaints from perpetrators. Their reporting experiences thus far gave them little confidence in platforms' willingness or ability to discern the truth.

\subsection{\textcolor{black}{Platform betrayal}} \label{conditions}
\color{black}
Betrayal presupposes a foundation of trust and dependence. We trace how victim-survivors initially expected platforms to keep them safe and how those expectations unraveled through repeated disappointments, \textcolor{black}{including platforms' failure to take proactive steps to prevent NCII (IBQ-2 Item 1) and the creation of online environments where NCII violations felt common, normal, and likely to occur again in the future (IBQ-2 Items 2 and 3)}. Repeated betrayal led some participants to abandon platforms altogether, feeling that they no longer ``belonged'' as members of the digital institution.

\subsubsection*{Platform dependence} 
Participants saw social media platforms as an integral part of their lives before and during their NCII experiences, supporting work, art, community building, political organizing, archiving, and fun. Participants who were victimized during the COVID-19 lockdown described the life-giving role that platforms played in their lives during this time. Especially for minors, platforms were often the main or only avenue for autonomy from their caregivers, including for sexual exploration, which fostered a heightened sense of dependence on platforms~\cite{gjika2023rape}. Even after this autonomy had been violated by and on platforms, younger participants still named platforms as the primary spaces where they could gain access to some semblance of independence, resulting in complicated dynamics of reliance. 

While experiencing NCII, participants developed a new sense of dependence on platforms, now as the only viable avenue that they felt could remedy the harm at hand. Many indicated that they were uncertain of whether and how they could report NCII to institutions other than platforms, particularly if their experience involved multiple jurisdictions or required the retrieval of digital evidence, such as the perpetrator’s DMs or deleted content. Several participants did attempt to report to traditional institutions such as their schools or law enforcement, but found that they appeared ignorant and ineffectual regarding online interventions. In particular, participants' experiences with traditional institutions showed that they were more focused on punishing perpetrators than removing or containing NCII. P8 relied heavily on online reporting after their attempts to report stolen Snapchat nudes to local law enforcement went ``horribly,'' with the officer indicating they were unable to assist. 

\begin{quote}
You could tell he was just like an old police officer who's used to traffic violations and doesn't know the real world and technology and all of the crime that can happen on technology \ldots he basically was like, well, you were the one that took [the photos] and put them on Snapchat.
\end{quote}

In other cases, participants felt online platform reporting was their only option because they were able to navigate it alone, without notifying---for instance---strict, religious parents. As with physical sexual assault, this preference became even more salient if disclosure of the harm would also have entailed other risky disclosures such as queer identity or stigmatized sexual interests~\cite{bedera2023could}. P1 saw platform reporting as her only real option, assuming it would reduce effort, limit harmful interactions, and provide the quickest intervention.

\begin{quote}
I don't know what my options are, my options are just to get this page taken down \ldots it felt hopeless, it really didn't feel like I had many options. So I just felt like I had to focus on Instagram and take it down from Instagram.
\end{quote}

\textcolor{black}{Without the cooperation of platforms, participants perceived they would be granted little access to justice, whether practical or epistemological. Participants' dependence on platforms as essential social spaces and, often, their only viable avenue for remedy intensified the stakes of platform responses and amplified their power to harm already vulnerable victim-survivors.}

\subsubsection*{Platform trust}
Before experiencing NCII, many participants described a sense of nonchalance or ``naivety'' associated with their online experiences. The internet was considered a place to play, experiment, and take risks, especially for young people~\cite{kender2025social,alluhidan2024teen}. P11 discussed their time online as a college student in the mid-2010s as embodying ``trust and safety'' in internet platforms, which rendered them comfortable enough to temporarily share a nude photo on their Tinder story: 

\begin{quote}
Sharing something and that it could show back up on the internet in another place just wasn't something that I was really conceiving of.
\end{quote}

Participants associated their beliefs about online protection and privacy with the assumption that platforms like Snapchat offered greater safeguards, given their branding as spaces intended for sensitive content. After initially experiencing NCII, participants expected platforms to act swiftly and comprehensively to remove content and punish perpetrator(s). These expectations were linked to the perceived power of the platforms, especially compared to what participants felt they could accomplish as individuals alone. Participants sometimes even assumed that platforms possessed powers beyond those of law enforcement, given platforms’ governance models and ownership of data. The process of handing reporting over to ``professionals'' and ``experts'' at platforms initially gave participants hope. P4 and P10 describe their initial expectations for Reddit and Instagram, respectively.

\begin{quote}
I was like, okay, just go to Reddit. They'll keep me safe. They'll deal with it. I just expected they'd be in communication with another corporation [or the] government to work on the legal stuff and the legal aspects, so I [can charge the perpetrator].
\end{quote}

\begin{quote}
When I first thought about reporting, I figured since Instagram \ldots was supposed to be one of the best, one of the more secure, but I thought they would have better IT and stuff, so I thought they would take care of it right away.
\end{quote}

\textcolor{black}{These beliefs in platform competence and goodwill demonstrate the trust participants placed in platforms as powerful digital institutions that they assumed would act to uphold their responsibilities and protect them as users. In fact, platform responses failed to align with these hopes. The dissonance between participants' trust and the realities of reporting created additional betrayals on top of the harms of NCII.}

\subsubsection*{Violated expectations and ``exiting'' the institution} 
Indeed, after experiencing reporting, participants described a strong sense of violated expectations, marked by frustration, shock, confusion, disappointment, and even hatred toward platforms and the corporations behind them. Many referenced a rupture in what they perceived as a contract between platforms and users, often articulated by terms of service that shaped their expectations of what platforms ``owe'' them~\cite{mcglynn_its_2021}. P3 articulated this sense of betrayal after realizing Instagram's privacy policy did not function as the protection she once assumed.

\begin{quote}
I think companies say, yeah, we're, we're creating this privacy policy to protect you. And then, now, I realized, oh, Instagram made it an absolute nightmare to try to get the information of someone who is deeply violating me. And you're doing it under the guise of protecting the users?
\end{quote}

Participants described an implicit social contract, in which platforms promised a sense of ``community'' while holding sole authority to enforce  safety. P4, who was threatened with the distribution of nude photos she shared as a minor on Discord, expressed surprise that reporting to the teenage moderator of her Discord server was more successful than reporting to Reddit. 

\begin{quote}
Within a few hours [of sharing screenshots to the teen moderator], the mod got him out. It was a pretty quick reporting process, which is really ironic because like how is this close-knit community of teens and their moderation quicker than these big corporations, and they were able to do it more effectively also?
\end{quote}

\textcolor{black}{In contrast to legal discourse that frames platform accountability narrowly, participants came to understand platforms as powerful institutions whose choices shaped the continuation of NCII and created new harms. They described new awareness of the power imbalance embedded in the platform-user relationship. This imbalance initially underpinned their trust and dependence but, through betrayal, came to feel withheld, misused, or weaponized. Violated expectations made platforms more visible to participants as institutions, not protective intermediaries but actors capable of enabling and exacerbating harm.}

This ruptured trust fundamentally altered how participants related to social media and their sense of belonging online. Many no longer felt welcome or safe after the combined injuries of NCII, burdensome reporting, and inadequate or harmful platform responses. Participants described losing access to once-valued spaces where they previously felt free to connect and express themselves. P2 vividly captured this loss: 

\begin{quote}
    I don't really post much at all anymore, and I used to love posting. I used to post my every thought! I used to be on there, very unhinged on my stories. I used to just have fun with it because it's fun at the end of the day. Social media is really fun, but I just kind of feel paranoid now. 
\end{quote}

For some, departure became not just a safety decision but a principled refusal to remain part of an institution that had betrayed them. After repeated failures by Instagram and Facebook, P5 chose to abandon Meta platforms entirely, explaining that she did not trust it had any intention to protect their users.

\begin{quote}
    And then I stopped using Meta this year just because fuck Meta, and I don't believe in anything they're doing anymore. So I don't use any of these accounts that I've poured my life into.
\end{quote} 

Participants' decisions to exit platforms reflected more than individual coping strategies. They mirror withdrawal patterns documented across the \ib literature. Exiting platforms was both a practical response and a recognition that institutions they depended on had misused their power, failed to uphold their commitments, and ultimately turned against them. 

\subsubsection*{Failure to prevent}
Participants' exits also revealed a deeper critique. Platforms not only mishandled reports but failed to prevent NCII altogether, allowing it to persist as an ordinary, foreseeable harm. Participants believed that platforms could have done more for them, listing specific ideas for online platform features that may have reasonably eliminated or reduced the harm they experienced. These included stronger user verification technologies, user IP bans, technologies to proactively scan for and remove NCII content, preemptive suggestions to block other users based on patterns of behavior, and digital hash technologies already used in other contexts. P5 believed these measures were basic, increasing her sense that platforms had abandoned their duties by failing to implement them. 

\begin{quote}
I came up with these lists of things I wanted [the platforms] to do, that I was like, they could be doing this, they could be doing this, they could be doing this, you know, and they weren't doing any of these things.
\end{quote}

Participants saw platforms’ failure to prevent harm as both omission and commission, noting that platforms intentionally did not adopt prevention tools because they had a profit motive to promote and retain abusive content. P5 described learning during her multi-year, ongoing attempt to curb an online harassment campaign using her NCII that Tumblr had implemented hash technologies for CSAM but not for content depicting adults. 

\begin{quote}
Why, why are we deciding that only child porn is a crime that we care about enforcing? Oh, I know, because we're adults, and so you can't, you can say it's consensual and then you can say oh, I'm going to make all this money off of this by leaving this post up. 
\end{quote}

On the other hand, participants believed that platforms actively prioritized protecting harmful content and its creators over supporting victimized users, allowing harm to spread where it could have been contained. P11 noted how their reporting experiences highlighted in new ways the ``explicit human design'' behind social media that allowed for the harm they experienced to occur in the first place. P3 believes that platforms ``intentionally created'' the barriers to reporting that \textcolor{black}{she} experienced. Through the reporting process, P10 recognized that platforms could have supported victim-survivors but chose not to, causing him to lose trust.

\begin{quote}
I know now not to put too much trust into these corporations because they're profit driven, and also their top priority isn't to keep the community safe. It's more just to make their money. I try to watch out more often and not to put too much trust in the whole reporting system because it's obviously flawed. 
\end{quote}

\textcolor{black}{Because participants viewed platforms as powerful institutions, they believed that meaningful prevention was both possible and reasonable to expect. They therefore interpreted platforms' failure to prevent NCII not as isolated error or unavoidable byproduct of complexity, but as evidence of institutional choices that designated victim-survivors as not worth protecting or, at least, less worth protecting than other users or the platform's own interests. The harm they experienced thus felt like acceptable collateral. Participants' own relative powerlessness to implement the protective measures they could clearly envision further amplified their sense of betrayal.}

\subsubsection*{Normalized culture of harm} 
Even participants' realizations that they were not alone in experiencing or reporting NCII contributed to a sense of powerlessness as the platform environment felt like one in which continued sexual harm was inevitable. Those who experienced this harm as minors described their NCII experiences as so common as to be the expected price of their engagement in sexual activity online. Looking back, P10 now describes the harms he experienced as a minor as ``just a natural part of growing up.''

Even before experiencing NCII, some participants assumed that the platform reporting process would be ineffective due to prior negative experiences with reporting other forms of harmful content. For example, P12 explained that their knowledge of Facebook’s ``really, really shitty reporting system'' for phishing scams and Instagram’s failure to remove racist comments led them to assume little would be done to curb NCII. Participants also noted the likelihood of experiencing NCII as explicitly correlated to the ``culture'' of various platforms, with some platforms fostering cultures where harms are more common. For instance, P10 named Discord as having a``shady'' culture, and P4 noted its many ``moderation fails.'' Similarly, P11 and P4 highlighted that Reddit's hands-off approach to sexualized content contributes to a higher perceived prevalence of harm. 

\textcolor{black}{Participants came to view platforms as sexually harmful environments, with differences between platforms serving as evidence of their active role in shaping or failing to curb these cultures. Repeated reporting disappointments, combined with the visible prevalence of abuse, reinforced a sense that NCII victimization is both common and likely. This recognition led some participants to forgo reporting altogether and instead adjust their own behavior, internalizing the notion that NCII is simply a normal part of online life.}

\color{black}

\subsection{What do some victim-survivors want instead?} \label{whatpeoplewant}

It is not victim-survivors' obligation to repair the systems that harmed them. Moreover, some of our participants felt they did not have the expertise required to weigh in on technical matters or expressed uncertainty about what would be best for others. Those who did offer suggestions said they would like more transparent communication from platforms, access to professionals to assist them, and technical features and systems to protect them from experiencing (more) harm. 

\subsubsection*{Transparent communication}
Participants reported a desire for platforms to be transparent about what they can or cannot do, even if their limitations are disappointing. P11 explains that if a platform cannot (or will not) act, it is better to receive confirmation of inaction and an explanation than to be left waiting for an indeterminate period or to receive no response at all. P8 elaborates: 

\begin{quote}
    I did like that I got the follow-ups from them, which was nice \ldots they'd be like, we didn't take down this website. So then I have to go back and find it. It was nice to actually physically see that some of those websites were taken down at the time. Always constantly getting that email confirmation of, you submitted it. You'll hear back from us. So I did appreciate that.
\end{quote}

Receiving clear communication from platforms enabled victim-survivors to promptly pursue other options, such as resubmitting reports with additional information or refocusing their efforts to different platforms. This communication thus supported victim-survivors' efforts to urgently enhance safety. Additionally, receiving responses from platforms increased participants' sense of agency. Knowing that their reports were moving through appropriate channels increased feelings of progress. 

Even beyond these practical concerns, participants expressed that they just want to be heard and to know that their reports---and efforts---matter. In speaking to platforms, P11 says
\begin{quote}
    Are you listening? Am I yelling into a void? I don't know. Is it just one human looking at my request when they have time and deciding that they agree with it? Is it a[n] algorithm that's weighing different information?
\end{quote}

The harm of inaction was compounded by platforms' refusal to explain their reasoning or, in some cases, to provide any response. Even a simple acknowledgment conveyed a recognition of victim-survivors' humanity and significantly improved their reporting experiences. Transparent communication, including rationale for inaction, is a key component of an ``adequate response'' (IBQ-2 Item 5) from an institution, signaling to the victim-survivor that they are valued users whose concerns have been, at least, considered. 

\subsubsection*{Access to someone (professional)}

Victim-survivors reported that they do not want to decipher the difficulties of reporting alone. Instead, they want teams of trained professionals assisting them and believe that assembling these teams should be the responsibility of platforms or government. P3 sums it up, ``I am tired, and I need someone who's an expert in this to do their job of what they're supposed to do.'' P10 wishes social media companies would invest in preventing abuse by ``[expanding] their support team so they have analysts whose main job would be to look through pictures and filter them out and figure out which ones are for photo shoots and which ones are intimate,'' making the parallel to IT-support. This need is especially urgent for minors, who may otherwise lack access to resources. P4 reflects, 

\begin{quote}
    I wish there was some anonymous support where I could tell someone about it and they could really deal with this. I wish there was another adult figure in my life that wasn't connected to my parents \ldots to help me navigate my emotions then and still stay safe online and also without having the threat of telling my parents and then having them criticize me.
\end{quote}

There is a steep learning curve for victim-survivors reporting NCII, as they must continually learn what their legal rights and options are, how each platform’s reporting system functions, and what steps they must take next. Yet this learning is required at a moment of intense isolation, fear, and exhaustion. Legal assistance and support networks are often too expensive to access, either financially or reputationally. Across interviews, participants emphasized their desire for a human presence to support them throughout the reporting process. For some, this support was imagined as emotional and cognitive, helping them articulate their needs and understand and process information influxes. For others, the most valuable form of support offloaded the labor of reporting, including proactive prevention tasks. Both forms of assistance could function as harm reduction by limiting opportunities for platforms to inflict further harm, reducing direct contact with victim-survivors and instead mediating interactions through supportive intermediaries. 

\subsubsection*{Features that prioritize their safety}

Participants consistently expressed frustration that platforms remain reactive rather than proactive and called for safety policies that intervene directly against perpetrators, rather than placing the burden on victim-survivors to justify their safety needs. P3 stressed \textcolor{black}{her} concern that platforms invoke data privacy as a reason for inaction. 
\begin{quote}
    I would want [Instagram] to understand the reality of what's happening on their platforms and the type of things that they are protecting, and I do think that consumer data is private and should be carefully dished out. But I wish that these corporations would consider all of the barriers that they have intentionally created, and how they are harming people.
\end{quote}

Other participants echoed this critique, emphasizing that platforms should prioritize protecting victims-survivors rather than scrutinizing their actions. P4 wished Discord would ``\ldots understand hey, deal with the perpetrator first before the victim. And like, get everything sorted out. Like, get [the perpetrator's] IP off the platform''. 

Building on their broader desire for human support, participants described wanted reporting features that would begin from a position of believing victim-survivors while still maintaining fairness and preventing abuse. P7, who used the DMCA process to pursue NCII removal, imagined processes focused primarily on establishing consent, while still including basic verification: 
\begin{quote}
Instead of trying to prove that the person doesn't own the work\ldots just proving that you're in the work, and it wasn't consensual. It should be enough\ldots It's mostly based on your word. I mean, obviously, [perpetrators] need to be able to have an opportunity to be heard and verify you're not lying because I don't want this process to be abusable, but generally speaking it would look pretty similar to the DMCA process\ldots just less focused on copyright and more focused on: this work was made, I'm in it, and either I don't want to be, or it wasn't supposed to be published. Effectively signing an affidavit to that effect\ldots
\end{quote}

After sorting through the backlog of messages to identify evidence most likely to be understood by moderators, P4 wished Discord would ``\ldots take a full DM into consideration instead of just specific messages.'' She also imagined automated systems on Reddit that could proactively detect potentially harmful DMs that contain words or messages known to be associated with NCII, either filtering them out or providing victim-survivors with warnings and resources.

Together, these reflections highlight participants' broader desire for technical systems that act meaningfully on information about perpetrators. Participants felt that current systems fail to hold perpetrators accountable, and they hoped for tools that send signals that victim-survivors' safety is valued and that their claims are believed, while also reducing the labor required to substantiate their reports. 

\color{black}

\section{Discussion} \label{discussion}

\subsection{Platforms and \ib}

Our data demonstrate that online platforms replicate many forms of \ib observed in traditional institutions. Participants emphasized their perspective that platforms hold exclusive authority over disciplining perpetrators and exercise unilateral power within their domain. Just as only an employer or university can remove a perpetrator from their ranks, participants saw platforms as the actors most capable of \textit{directly} stopping abuse online. Through reporting, participants became increasingly aware of platforms' power, just as they saw it turned against them. Their accounts reflect the logic of \ib, in which victim-survivors must depend on the very institution that fails to protect them~\cite{bedera2024wrong}. 

The \textit{online} nature of platforms, however, introduced dynamics that make \ib qualitatively different for NCII victim-survivors. While digital environments replicate both offline social and institutional dynamics, participants' experiences highlight how platform affordances create novel harms~\cite{walther1996computer, brown2022problem}. For instance, the removal of an online perpetrator, even via another institution (such as a prison), does not end the violation because NCII exists beyond the bounds of space and time. Participants described NCII being replicated across websites, amplified by algorithms, turned into deepfakes, and resurfacing indefinitely~\cite{mcglynn_its_2021, compton2024deepfake}. When platforms (or rather, the internet) failed our participants, ``departing'' the institution altogether was made impossible by the ubiquity, timelessness, and spaceless-ness of the web. 

The platform's distinct role becomes even more visible in participants' attempts to remove abusive content. The scale, spread, and speed that define the internet fundamentally shaped how betrayal unfolded. Unlike any singular traditional institution, platforms such as Instagram govern billions of users and rely heavily on automated systems to do governance work ~\cite{gillespie_content_2020,donovan2018algorithmic}.
As Citron warns, automated decision-making risks ``dismantling critical procedural safeguards at the foundation of administrative law \ldots [seamlessly combining] rule-making and individual adjudications without the critical procedural protections''~\cite{citron2007technological}. 

Traditional institutions such as universities or the military operate within legal frameworks that include \textit{some} external accountability measures, even if their efficacy is limited as documented in \ib literature. Meanwhile, participants encountered platforms whose accountability structures were largely self-defined and unenforceable. Content removal processes were dictated by platforms' individual Terms of Service, with minimal external oversight, often alongside claims of exemption from existing legal frameworks~\cite{caplan2016controls,gillespie2010politics}. Beyond narrow avenues like the DMCA---which victim-survivors often found challenging or ineffective---participants had no realistic mechanisms to challenge a platform's decisions ~\cite{qiwei2024reporting,qiwei2024law}. As a result, platforms acted simultaneously as crime scene, judge, and jury, inventing the rules, enforcing them internally, and allowing no meaningful right of appeal~\cite{gillespie2010politics,gillespie2018custodians,gorwa2019platform}. 

Ultimately, participants' experiences highlight a central precondition of \ib: dependence. While traditional institutions carry legal or moral duties toward those who rely on them, even the term ``platform'' cultivates the appearance of neutrality to avoid responsibility for user harm~\cite{bowman2021public,gillespie2010politics, halman1994religion,pollman2019corporate}. This positioning frees platforms from regulatory pressure while minimizing their obligations, even when their actions resemble forms of institutional betrayal that would incur liability elsewhere. Participants varied in their initial trust---some expected careful handling, others anticipated dismissal---but all reported because they perceived few, if any, viable paths to relief. This lack of agency forced participants into another abusive power dynamic with the institution or, alternately, produced forced marginalization and exclusion. 

\color{black}

\subsection{Building trauma-informed reporting systems}

Trauma-informed systems are necessary to counter \ib ~\cite{chen2022trauma,scott2023trauma,smith_dangerous_2013}. The results of this study show that NCII reporting systems are currently not trauma-informed. Below, we outline directions for new technologies, design, and policy changes. 

Automation was a connecting thread through our interviews. Participants wished both for \textit{more} human involvement and \textit{more} automation to support them in reporting. Specifically, they suggested \textit{decisions} about whether and how to report and the provision of emotional reassurance during the process should be supported by humans, while \textit{procedural} tasks such as locating forms on a website should be automated. One key direction for HCI research is to explore when and how to redirect to humans, and how the user, human, and system interact. 

Many of the suggestions participants made are already fully achievable and mirror conveniences already commonplace in other online processes. Participants were especially frustrated by having to return to the same sites each time a new post appeared. Other domains have solved this problem long ago. A job applicant can submit one resume to multiple jobs with a single click. College applicants may use the Common App for multiple university applications. Victim-survivors deserve at least the same level of ease for repetitive tasks that are emotionally costly and retraumatizing. In a similar vein, participants described how reporting interfaces restricted their ability to express the full extent of harm incurred. This left them feeling unheard, deepening their victimization. One alternative is to build intake systems that allow fuller initial storytelling and then route users to the relevant rules. Such systems could also contain cultural context in ways for which current moderation practices rarely account ~\cite{batool_expanding_2024,schoenebeck2023online}.

Participants say they need access to someone professional for dealing with NCII, which raises questions about who should fill that role. If tech companies continue to fall short and benefit from the conditions that allow NCII to circulate, they should ideally also fund the structures that compensate for those failures. We may look to rape crisis centers with established victim advocacy programs as models for what NCII support structures could look like, though the scale of the problem may require an entirely new institution~\cite{campbell2006rape,bedera2022illusion}. Implementing these models would require policy changes. NCII victim-survivors consistently emphasize that meaningful redress requires systemic, coordinated, and survivor-centered action in the U.S. and beyond~\cite{eaton_victim-survivors_2024,flynn2025sexualized}. In the U.S., they call for a national framework led by an independent Office for Online Safety, similar to Australia’s eSafety Commissioner or New Zealand’s NetSafe, to unify fragmented efforts, fund survivor services, and run proactive public education. Survivors stress that stronger national laws and cross-state coordination efforts are essential, including comprehensive statutes that cover threats and deepfake creation, along with simplified legal processes and advocacy support to help victims navigate them~\cite{flynn2025sexualized}. Together, these recommendations underscore that addressing NCII requires a multi-layered strategy, uniting technology, law, and survivor-led systems of care. 

\color{black}

\subsection{Implications for addressing NCII}

Our findings show that part of the overall psychological toll of NCII can be traced directly to how platforms design, process, and respond to reports. \textcolor{black}{Survivors' perceptions of their victimization are inseparable from their experiences with reporting.} Naming \ib elucidates how online systems are active contributors to harm and repositions platforms as accountable institutions rather than neutral infrastructures. Platforms, like other institutions, cannot simply stand behind Section 230 protections as the law has thus far allowed them to do \cite{citron2018problem}. Instead, they carry obligations for care and protection \cite{citron2023fix}. Because victim-survivors perceive these obligations, continuing to flout them will result in further harm. Recognizing this expands accountability beyond individual NCII perpetrators and distributors to include the companies that structure reporting and remediation.

Second, naming \ib in platforms creates new levers for intervention. Much of the existing literature, policy, and technical work on NCII documents the harms perpetrators exert on victim-survivors \cite{mcglynn_beyond_2017,umbach_prevalence_2025,mcglynn_its_2021}. In contrast, \ib identifies harms that arise specifically from the institution itself. This distinction allows us to parse which harms are addressable at the platform level and to create actionable intervention points where few currently exist. If we cannot eliminate all perpetrators or prevent all distribution of NCII, we can at least reduce the burden platforms impose on victim-survivors by building reporting systems that are truly accessible, consistent, effective, and humane, restoring some sense of safety and agency while avoiding (re)traumatization. 

Third, lessons can be drawn from institutions that have already confronted the consequences of betrayal. For decades, the U.S. military allowed commanders to control sexual assault prosecutions, leading to retaliation and inadequate outcomes~\cite{andresen2019institutional}. In 2023, this authority was removed from commanders and given to newly created Offices of Special Trial Counsel, independent military prosecutors outside the chain of command with the power to pursue sexual assault and other serious crimes~\cite{armyOSTC}. \textcolor{black}{Additional examples of steps institutions and policymakers can take to counter \ib can be found through the Center for Institutional Courage's guidelines~\footnote{https://www.institutionalcourage.org/resources-for-changemakers}}
. Online platforms currently hold similar unilateral power over NCII reporting. Like the military, platforms should be checked by independent bodies empowered to investigate and enforce takedowns on behalf of victim-survivors, rather than leaving the work in the hands of the same systems that enabled the harm.

Finally, our findings imply policy consequences. While the new TAKE IT DOWN Act mandates that NCII be removed within 48 hours of reporting, it does not specify how this goal ought to be implemented or enforced~\cite{grimmelmann2025deconstructing,TAKEITDOWN_Act_2025}. In practice, this means that platforms will decide on procedures, thresholds, and exceptions around what is considered a valid report. Platforms will still write the rules, adjudicate the violations, and execute decisions with little external oversight \cite{gillespie2018custodians,caplan2023networked}. The burden of reporting each instance of NCII also still falls to victim-survivors themselves. Future legislation must set minimum standards for reporting pathways, enforce consistency across platforms, and ensure that victim-survivors encounter transparency and care at each step. A centralized removal mechanism is needed to restore trust and reduce harm.

\section{Conclusion} \label{conclusion}

Our study reveals how victim-survivors navigate platforms when reporting NCII and how those encounters are shaped by the platforms themselves. Naming and mapping institutional betrayal in the online platform context creates new points of leverage. Through institutional betrayal theory, we generate actionable implications for policy and design, opening novel pathways for building reporting systems that reduce NCII harms. Our findings illuminate how platforms betray users, raising urgent questions of accountability, governance, and design. 

\bibliographystyle{ACM-Reference-Format}
\bibliography{1_sample-base}

\end{document}

%% file: 1_figures.tex
\usepackage{array,booktabs,tabularx,caption, float} 

\newcommand{\bexamples}{
\begin{table}[t]
\centering
\small
\setlength{\tabcolsep}{5pt}%
\renewcommand{\arraystretch}{1.2} 
\begin{tabular}{@{}  
  c    
  p{4cm}   
  p{4.5cm}    
  p{4.5cm}  
@{}}
\toprule
\textbf{IBQ-2} & \textbf{Institutional betrayal} & \textbf{Applied to online platforms} & \textbf{NCII} 
\\
\midrule

1 & Not taking proactive steps to prevent harm 
& Platform fails to implement reasonable safeguards or proactive measures that could have reduced risk to users
& Not implementing hash-mapping, not training moderators, not banning certain subreddits 
\\

2 & Creating environment where experience seemed normal
& Platform norms, culture, or design features make harmful practices appear acceptable or trivial
& Minimal visible enforcement, rules are hidden and buried, jokes or memes about this harm are trivialized 
\\

3 & Creating environment where experience seemed more likely to occur 
& Platform structures or algorithms amplify risks
& Recommendation algorithms promoting non consensual content, lack of account verification
\\

4 & Making it difficult to report experience 
& Pathways to surface concerns are confusing, inaccessible, or ill-fitted
& Reporting interface is confusing, requires a lot of steps, lack of clear contact pathways
\\

5 & Responding inadequately to experience, if reported 
& Platform replies are delayed, generic, opaque, or insufficient to meaningfully address the harm
& Platform replies with generic messages, delays action, doesn't provide information on outcomes
\\

6 & Mishandling the case 
& Platform processes or enforcement actions worsen the situation, creating new risks
& Allows the abuser back on platform? when abuser is told who the report came from (DMCA)
\\

7 & Covering up experience
& Platform obscures or conceals the scope of harm, through silence, under-reporting, or lack of transparency
& Moderators don't address the bigger underlying problem, incidents are not put in transparency reports 
\\

8 & Denying experience 
& Platform dismisses user concerns, claiming no violation occurred or refusing to acknowledge evidence
& Platform responds that no harm was found despite evidence
\\

9 & Punishing the victim-survivor 
& Users who raise concerns face retaliation, penalties, or negative consequences
& Getting account banned for reporting, (fear of) counter complaint
\\

10 & Suggesting experience might affect reputation of institution 
& Platform discourages disclosure or critique to protect its brand
& 
\\

11 & Creating environment where one no longer felt like a valued member of the institution 
& Platform’s neglect or responses erode user belonging
&  
\\

12 & Creating environment where continued membership is difficult 
& Users disengage or leave the platform entirely, unable to participate under harmful conditions
& Leaving the platform, blocking people, leaving certain online spaces 

\\

\bottomrule
\end{tabular}
\caption{The institutional betrayal questionnaire version 2 (IBQ-2) \cite{smith2017insult}, the IBQ-2 applied to online platforms, and how betrayal may manifest in non-consensual intimate media reporting online.}
\label{tab:betrayal}
\end{table}
}

\newcommand{\reports}{

\begin{table}[t]
\centering
\small
\setlength{\tabcolsep}{4pt}%
\renewcommand{\arraystretch}{1.2} 
\begin{tabular}{@{}  
  c    
  p{3.5cm}    
  c    
  c    
  p{4.5cm}  
  c    
  c    
@{}}

\toprule
\textbf{P\#} &\textbf{Demographics} & \textbf{Year} & \textbf{Was minor?} & \textbf{Platforms, sites, \& institutions involved} & \textbf{Ongoing} & \textbf{Time to ``resolve''}\\
\midrule
1 & Black/African-American \& Middle Eastern/North African cis woman, bisexual, pansexual
 & 2020 & No & WhatsApp, Twitter, Instagram & No & 1 year \\
2 & Black/African-American \& Hispanic/Latinx non-binary person, bisexual &2018 & Yes & Instagram, Snapchat, school, LE, child welfare system, and FBI & No & 6-7 months\\
3 & \textcolor{black}{White cis woman} &2023 & No & Telegram, 4chan, Reddit, Instagram, facial search sites, Google, Bing, pornographic sites, LE, FBI, civil court system & Yes & --\\
4 & Black/African-American cis woman, queer &2024 & Yes & Discord, Reddit & No & Several months\\
5 & White cis woman, queer, bisexual, pansexual
 &2018 & No & Tumblr, Snapchat, Twitter, Facebook, pornographic sites, workplace, LE, civil court system & No & --\\
6 & Hispanic/Latinx cis woman, queer &2011 & Yes & Facebook & No & Several hours \\
7 & White trans woman, asexual, lesbian &2023 & No & Pornographic sites, FetLife, civil court system & No & 1 year \\
8 & -- &2024 & No & Google, Snapchat, pornographic sites, FTC, LE, FBI, workplace & Yes & --\\
9 & -- &2012 & No & Pornographic sites, LE & No & 6 months\\
10 & Asian cis man, bisexual &2022 & No & Instagram, Discord, StopNCII & No & 2-3 years\\
11 & White non-binary person, queer &2016, 2022 & No & Reddit, Instagram, Tinder, Facebook, website hosting services & No & Several months\\
12 & Black/African American, American Indian/Alaska Native, \& Hispanic/Latinx trans, non-binary person, queer
 &2013 & Yes & KiK, photo sharing sites, YouTube & No & 12 hours\\
13 & White cis man, bisexual &2018 & Yes & Facebook, GIF sharing sites & No & Several hours\\

\bottomrule
\end{tabular}
\caption{Table summarizing participants' interactions with online platforms and other institutions while reporting non-consensual intimate media. Some participants elected not to disclose demographic data or to disclose only certain data; as a result, we share partial or no information for several participants. Throughout the text "they/them" pronouns are used to refer to participants who did not disclose their preferred pronouns, in addition to participants who use "they/them" pronouns. \textit{Year} refers to the year the participant became aware of or involved in the main NCII incident. Five participants were minors at the time of the incident. \textit{Ongoing} means the participant is still actively engaged in reporting their own abusive content online. Time to \textit{resolution} is when the participant described that they could stop participating in the reporting process. It does not necessarily represent the timeline for removal of all NCII content, particularly given that some participants actively avoid gaining awareness of new or re-posted content. \textit{Time to ``resolve''} only applies to cases that are no longer ongoing. \textit{LE} refers to law enforcement.}

\label{tab:reports}
\end{table}
}